\documentclass[11pt]{article}
\usepackage[utf8]{inputenc}
\usepackage[authoryear]{natbib}
\usepackage{bibunits}
\defaultbibliographystyle{apalike}
\defaultbibliography{ref}
\usepackage[a4paper]{geometry}
\usepackage[title]{appendix}
\usepackage{xcolor}
\usepackage{setspace}
\usepackage{algorithm}
\usepackage{algpseudocode}
\algrenewcommand\algorithmicrequire{\textbf{Input:}}
\algrenewcommand\algorithmicensure{\textbf{Output:}}

\usepackage{amsmath}

\usepackage{amsfonts}
\usepackage{bbm}
\usepackage{graphicx}
\usepackage{float}
\usepackage{booktabs, multirow}

\usepackage{makecell}
\usepackage{amsthm}
\theoremstyle{definition}

\usepackage{url}
\usepackage{comment}
\usepackage{subcaption}
\numberwithin{equation}{section}

\title{Spectral subsampling MCMC for L{\'e}vy-driven continuous-time ARMA models with expensive likelihood contributions}
\author{Thomas Goodwin\thanks{Human Technology Institute, University of Technology Sydney.} \and Matias Quiroz\thanks{School of Mathematical and Physical Sciences, University of Technology Sydney, Sydney, NSW, Australia.}$\;\,^*$ \and Robert Kohn\thanks{School of Economics, University of New South Wales, Sydney, NSW, Australia.} \and Feng Li\thanks{Guanghua School of Management, Peking University, Beijing, China. Corresponding author email: feng.li@gsm.pku.edu.cn.}}
\date{}

\begin{document}
\maketitle

\begin{abstract}
Subsampling-based Markov chain Monte Carlo (MCMC) algorithms aim to accelerate Bayesian inference by evaluating the likelihood using only a subset of the data at each iteration. However, in many standard tall-data applications, individual likelihood contributions are inexpensive to evaluate and the resulting reductions in actual computing time are often substantially smaller than the nominal reduction in data size due to computational overhead. We study a different computational regime arising in frequency-domain inference for continuous-time processes observed at equally spaced discrete time points. This gives rise to aliasing, whereby each contribution to the Whittle likelihood requires summation over shifted frequency components, unlike standard discrete-time spectral settings where spectral evaluations do not require such summation. We demonstrate that this structure makes subsampling MCMC, a subsampling-based MCMC approach that estimates the log-likelihood using data subsampling and efficient control variates, particularly effective for reducing computational cost. We illustrate the approach for Bayesian frequency-domain inference in discretely observed continuous-time autoregressive moving average models driven by finite second-moment L{\'e}vy processes.
 
\end{abstract}

\section{Introduction}

Scalable Bayesian inference remains challenging in models where likelihood evaluations are computationally intensive; see \cite{liu2001monte} for an overview of Monte Carlo methods in scientific computing. Subsampling-based Markov chain Monte Carlo (MCMC) methods aim to reduce this cost by estimating the likelihood using only a subset of the data at each iteration; see \cite{bardenet2017markov} for an excellent review of subsampling methods for MCMC. While these methods have been widely studied in settings where the likelihood decomposes into a large number of independent and inexpensive contributions, they do not always lead to substantial reductions in actual computing time. This setting, which we refer to as the tall-data, inexpensive-likelihood-contributions regime, is common in applications such as generalised linear models, where individual likelihood contributions can be computed efficiently using vectorised operations. Although subsampling reduces the number of terms in the likelihood, it introduces additional overhead through data access and indexing, and the resulting speed-ups are often far smaller than what would be suggested by the reduction in data size. Consequently, subsampling in such settings typically leads to much smaller computational gains than suggested by the reduction in sample size. For this reason, performance is often reported in terms of the fraction of the data used, which can be interpreted as an implementation-free proxy for computational cost \citep{maclaurin2014firefly, liuexact2015fireflycousin, korattikara2014austerity, bardenet2017markov, quiroz2019speeding, quiroz2021block, salomone2020spectral, villani2024spectral}. Notable exceptions arise in models with recursive likelihoods, where each likelihood evaluation involves sequential computation over the full dataset, and recent subsampling approaches have been proposed to accelerate these computations \citep{aicher2025stochastic,quiroz2026stabilised}.

Our paper considers a different computational regime arising in frequency-domain inference for continuous-time models. In frequency-domain inference, likelihood-based methods are commonly implemented using the Whittle likelihood \citep{whittle1953estimation}, which provides an asymptotic approximation to the corresponding time-domain likelihood \citep{kent1996spectral} and is evaluated over a set of frequency components that form the individual contributions to the likelihood. In this setting, these contributions require computing an aliased spectral density, which arises when a continuous-time process is observed at discrete times and combines information across multiple frequency components. Consequently, each contribution has non-negligible computational cost, and the overall computational burden is determined by both the number of contributions and the cost of evaluating each one. This structure gives rise to a setting in which subsampling can directly reduce computational cost. By evaluating the likelihood at a subset of frequencies, the number of expensive spectral evaluations is reduced, leading to tangible gains in computational cost. The present work therefore studies a class of models in which subsampling MCMC yields genuine computational speed-ups.

We consider likelihood-based inference for continuous-time autoregressive moving average (CARMA) models driven by finite second-moment L{\'e}vy processes, observed at equally spaced discrete time points. The L\'{e}vy-driven formulation extends the Brownian-motion-driven Gaussian setting to a broader class of processes capable of exhibiting features such as jumps, asymmetry, and heavy-tailed behaviour. These models provide a flexible framework for representing systems whose latent dynamics evolve in continuous time while observations are recorded discretely, with applications in econometrics \citep{bergstrom1988history,brockwell2011estimation}, control theory \citep{gillberg2009frequency}, and engineering \citep{mossberg2004fast}. Continuous-time formulations can also naturally accommodate multivariate settings where different time series may be observed at different regular sampling frequencies. An attractive feature of frequency-domain inference for finite second-moment L\'{e}vy-driven CARMA models is that the Whittle likelihood depends only on the second-order structure of the process through its spectral density. Consequently, the same frequency-domain methodology applies across a broad class of such models, despite potentially markedly different time-domain dynamics, without requiring separate time-domain likelihood derivations for each model. Since the Whittle likelihood depends only on the second-order structure of the process, inference in this framework does not attempt to identify the full set of L\'{e}vy parameters. A computational complication, however, is that the spectral density of the observed process is an aliased version of the underlying spectrum, which leads to likelihood contributions that are more expensive to evaluate than in standard discrete-time spectral settings. This structure makes CARMA models a natural setting for subsampling-based inference, as the computational cost is driven by evaluating the individual spectral contributions.

An alternative approach in frequency-domain inference is the debiased Whittle likelihood \citep{sykulski2019debiased}, which reduces the small-sample bias of the standard Whittle likelihood by replacing the spectral density with the expected value of the periodogram. This modification corrects for finite-sample effects, including spectral blurring and, for discretely observed continuous-time processes, aliasing \citep{sykulski2019debiased}. See \cite{guillaumin2022debiased} for an extension to spatial data and \cite{goodwin2025calibrated} for its application in Bayesian inference. The debiased Whittle likelihood has the same asymptotic computational order as the standard Whittle likelihood, but results in a likelihood that is less amenable to subsampling strategies designed to reduce computational cost relative to the full-data likelihood. Section \ref{subsec:computational_regimes} characterises regimes in which a subsampling-based implementation of the standard Whittle likelihood can be computationally more efficient than both its full-data counterpart and the debiased Whittle likelihood.

Our main contribution is to demonstrate that subsampling MCMC methods can yield genuine reductions in computational cost in settings where likelihood contributions are individually expensive to evaluate. Focusing on frequency-domain inference for continuous-time models observed at equally spaced discrete time points, we show how subsampling can be used to reduce the cost of evaluating aliased spectral densities while preserving the form of the Whittle likelihood and enabling unbiased log-likelihood estimation within a subsampling framework. In addition, we provide a simple characterisation of regimes in which subsampling-based implementations of the Whittle likelihood can be computationally more efficient than both their full-data counterparts and alternative approaches such as the debiased Whittle likelihood. 

The rest of the paper is organised as follows. Section \ref{sec:methodology} briefly presents the methodology used in the paper, including frequency-domain inference, subsampling MCMC, and our characterisation of the computational regime in which genuine central processing unit time gains are achievable. Section \ref{sec:experiments} contains simulation experiments validating the Whittle likelihood assumptions for three CARMA models, assessing the accuracy of the full-data Whittle posterior relative to the time-domain posterior, the accuracy of the subsampled Whittle posterior relative to the full-data Whittle posterior, and the effect of aliasing truncation. Section \ref{sec:application} provides a real-data application to high-frequency bitcoin data. Section \ref{sec:conclusion_and_future_research} concludes and outlines future research. Finally, Appendix \ref{app:carma_time_domain} briefly outlines the time-domain formulation of the continuous-time ARMA models considered in the paper.

\section{Methodology}\label{sec:methodology}

\subsection{Frequency-domain inference}\label{eq:model_and_likelihood}

We consider likelihood-based inference for continuous-time stationary processes 
$\{Y(t)\}_{t \geq 0}$  with parameter vector $\boldsymbol{\theta}$, observed at regularly spaced times $s\delta$ for $s = 0, \dots, N-1$, where $\delta > 0$ is the sampling interval and $N$ is the sample size. The resulting discrete-time observations are
$$y_s^{(\delta)} = Y(s\delta), \,\, s = 0, \dots, N-1.$$
For notational simplicity, we assume throughout that $N$ is odd, although the extension to even $N$ is straightforward. 

In frequency-domain inference, parameter estimation is commonly based on the Whittle log-likelihood \citep{whittle1953estimation}, which we simply refer to as the Whittle likelihood. This likelihood is obtained from the discrete Fourier transform of the observations. Let
\begin{align} \label{eq:DFT}
J_\delta(\omega_k)
=
\sqrt{\frac{\delta}{N}}
\sum_{s=0}^{N-1}
y^{(\delta)}_s e^{-\mathrm{i}\omega_k s\delta}
\end{align}
denote the discrete Fourier transform (DFT) at the Fourier frequencies
$$\omega_k
=
\frac{2\pi k}{\delta N},
\,\,
k = 1, \dots, \lfloor N/2 \rfloor.$$
The corresponding periodogram ordinates are defined as 
\begin{align*}
\mathcal{I}_\delta(\omega_k)
=
|J_\delta(\omega_k)|^2.
\end{align*}
Under standard regularity conditions (see, for example, \citeauthor{gelfand2010handbook}, \citeyear{gelfand2010handbook}), the periodogram ordinates are asymptotically independent with
\begin{align}\label{eq:periodogram_dist}
\mathcal{I}_\delta(\omega_k) \sim \mathrm{Exp}\big(f_\delta(\omega_k; \boldsymbol{\theta})\big),
\end{align}
where $f_\delta(\omega; \boldsymbol{\theta})$ is the spectral density of the observed discrete-time process and  $\mathrm{Exp}(\mu)$ denotes the exponential distribution with mean $\mu$. This leads to the Whittle likelihood
\begin{equation} \label{eq: carma whittle log likelihood}
  \ell_{\text{W}}(\boldsymbol{\theta}) =  -\sum_{k=1}^{(N-1)/2}  \left(\log f_{\delta}(\omega_k; \boldsymbol{\theta}) + \frac{\mathcal{I}_{\delta}(\omega_k)}{f_{\delta}(\omega_k; \boldsymbol{\theta})} \right),
\end{equation}
which is an asymptotic approximation to the exact time-domain likelihood
\begin{align*}
    \ell(\boldsymbol{\theta}) = \log p(\mathbf{y}|\boldsymbol{\theta}), \,\, \mathbf{y} = (y_0^{(\delta)},\dots,y_{N-1}^{(\delta)}),
\end{align*}
for stationary processes \citep{kent1996spectral}. A key advantage of the Whittle likelihood is that it can be computed in $\mathcal{O}(N\log N)$ operations using the fast Fourier transform (FFT) \citep{cooley1965algorithm}, whereas evaluating the corresponding time-domain likelihood may require up to $\mathcal{O}(N^2)$ operations for stationary processes.

For discretely observed continuous-time processes, the spectral density $f_{\delta}(\omega_k; \boldsymbol{\theta})$ in \eqref{eq: carma whittle log likelihood} is obtained by evaluating the aliased spectrum
\begin{align}\label{eq:aliased_spectrum}
f_\delta(\omega; \boldsymbol{\theta})
= \sum_{j=-\infty}^{\infty}
f\left(\omega + \frac{2\pi j}{\delta}; \boldsymbol{\theta}\right), \,\, \omega \in \left[-\frac{\pi}{\delta},\frac{\pi}{\delta}\right],
\end{align}
at the Fourier frequencies $\omega_k$, where $f(\cdot; \boldsymbol{\theta})$ denotes the spectral density of the underlying continuous-time process, defined for $\omega\in\mathbb{R}$. This expression arises from aliasing, whereby equally spaced discrete-time observations cannot distinguish between frequencies that differ by integer multiples of $2\pi/\delta$ due to the periodicity of the Fourier basis functions. In practice, \eqref{eq:aliased_spectrum} is approximated by a finite sum, truncating the contributions to $|j|\leq K$, where $K$ controls the accuracy of the approximation. Evaluating $f_\delta(\omega; \boldsymbol{\theta})$ therefore requires $\mathcal{O}(K)$ operations for each frequency. As a result, evaluating the Whittle likelihood in \eqref{eq: carma whittle log likelihood} requires $\mathcal{O}(K N)$, assuming the periodogram is computed once at a cost of $\mathcal{O}(N \log N)$ operations. Since the periodogram does not depend on $\boldsymbol{\theta}$, it does not need to be recomputed during an optimisation or simulation algorithm. A key feature of this setting is that each likelihood contribution has non-negligible computational cost when $K$ is moderate to large, as it requires evaluating the aliased spectral density. This motivates the use of subsampling methods that reduce the number of likelihood contributions evaluated at each iteration.

\subsection{Subsampling Markov chain Monte Carlo}

We consider the subsampling Markov chain Monte Carlo (MCMC) framework developed in \cite{quiroz2019speeding}; see \cite{quiroz2018subsampling, quiroz2023wiley} for introductory surveys. Several extensions and applications of this approach have been proposed, including \cite{quiroz2021block, dang2019hamiltonian, quiroz2018delay}. In particular, \cite{salomone2020spectral} applies this framework to Whittle likelihoods for discrete-time time series models; see \cite{villani2024spectral} for an extension to multivariate time series. Following \cite{salomone2020spectral}, we refer to the frequency-domain application of subsampling MCMC as spectral subsampling MCMC, although for brevity we write subsampling MCMC in the paper.

Let $n=(N-1)/2$ denote the number of frequency contributions in \eqref{eq: carma whittle log likelihood}. Then the Whittle likelihood in \eqref{eq: carma whittle log likelihood} can be written as
\begin{align} \label{eq:Whittle_decomposed}
    \ell_{\text{W}}(\boldsymbol{\theta}) = \sum_{k=1}^{n} \ell_k(\boldsymbol{\theta}),
\end{align}
where each term $\ell_k(\boldsymbol{\theta})$ depends on $\mathcal{I}_{\delta}(\omega_k)$ and $f_{\delta}(\omega_k; \boldsymbol{\theta})$. The additive structure in \eqref{eq:Whittle_decomposed}, together with the fact that each $\ell_k(\boldsymbol{\theta})$ depends only on the corresponding frequency $\omega_k$, allows the log-likelihood to be estimated using a random subset of frequency contributions at reduced computational cost. \cite{quiroz2019speeding} employ the difference estimator combined with uniform subsampling to construct an unbiased estimator of the log-likelihood. Specifically, let $u_1,\dots,u_m$, where $m$ is the subsample size, denote indices sampled independently and uniformly from $\{1, \dots, n\}$. The difference estimator takes the form
\begin{align}\label{eq:diff_estimator}
    \widehat{\ell}_{\text{W}}(\boldsymbol{\theta})=\sum_{k=1}^n q_k(\boldsymbol{\theta}) + \frac{n}{m}\sum_{i=1}^m \left(\ell_{u_i}(\boldsymbol{\theta}) - q_{u_i}(\boldsymbol{\theta}) \right),
\end{align}
where $q_k(\boldsymbol{\theta})$ are control variates that approximate $\ell_k(\boldsymbol{\theta})$. We use quadratic control variates based on a Taylor expansion around a reference parameter value $\boldsymbol{\theta}^\star$, following \cite{bardenet2017markov,quiroz2019speeding}. This allows the first term in \eqref{eq:diff_estimator} to be precomputed once at a cost of $\mathcal{O}(n)$ operations, after which it can be computed at $\mathcal{O}(1)$ during the Markov chain Monte Carlo iterations. 

The log-likelihood estimator in \eqref{eq:diff_estimator} can be exponentiated to estimate the likelihood $\exp(\ell_W(\boldsymbol{\theta}))$ via  $\exp(\widehat\ell_W(\boldsymbol{\theta}))$. However, by Jensen's inequality, this estimator is biased. \cite{ceperley1999penalty} note that the bias can be removed under idealised assumptions, and \cite{quiroz2019speeding} incorporate the resulting estimator into a pseudo-marginal framework \citep{beaumont2003estimation, andrieu2009pseudo}. Under substantially weaker assumptions, the likelihood estimator is generally biased, but the resulting pseudo-marginal algorithm targets a perturbed posterior that is typically very close to the true posterior; see \cite{quiroz2019speeding} for details.

\cite{johndrow2020no} show that subsampling-based Markov chain Monte Carlo methods are generally limited in their ability to scale. However, they identify efficient control variates as a setting in which meaningful speed-ups may be achievable; see also \cite{rudolf2026perturbations}. In the present setting, the individual contributions $\ell_k(\boldsymbol{\theta})$ admit accurate quadratic approximations after grouping (see Section \ref{subsec:posterior_accuracy_experiment}), enabling the construction of efficient control variates and supporting the practical effectiveness of subsampling.

\subsection{Computational regimes}\label{subsec:computational_regimes}

To obtain meaningful reductions in central processing unit (CPU) time, the effectiveness of subsampling methods depends critically on the computational cost of individual likelihood contributions. Let $n$ denote the number of log-likelihood contributions in \eqref{eq:Whittle_decomposed} and $C$ denote the cost of evaluating a single contribution $\ell_k(\boldsymbol{\theta})$. Then evaluating the full likelihood requires $\mathcal{O}(nC)$ operations, while subsampling-based estimators require $\mathcal{O}(mC)$, where $m$ denotes the subsample size.

In the present setting, each contribution $\ell_k(\boldsymbol{\theta})$ requires evaluating the aliased spectral density in \eqref{eq:aliased_spectrum}, which, when truncated to $K$ terms, requires $\mathcal{O}(K)$ operations. Hence, the cost per contribution satisfies $C=\mathcal{O}(K)$. In addition, the use of control variates introduces an initial computational cost to construct the approximations $q_k(\boldsymbol{\theta}^\star)$ for all $k = 1, \dots,n$. However, this cost is incurred only once and can be amortised over subsequent iterations of the algorithm. In contrast, the cost of evaluating likelihood contributions and their approximations for the subsample arises at every iteration. Hence, the computational savings from subsampling are determined primarily by the per-iteration cost, rather than the one-off cost of initialising the control variates. Subsampling yields meaningful reductions in CPU time only when the cost per contribution is sufficiently large. In particular, subsampling is effective when $n\gg m$ and the per-iteration cost satisfies $$\mathcal{O}(mC) \ll \mathcal{O}(nC).$$
When likelihood contributions are inexpensive to evaluate, the reduction from $n$ to $m$ may involve overhead (e.g. data indexing required by subsampling), preventing the realised CPU reduction from reflecting the nominal ratio $n/m$. This occurs when the overhead cost is of the same order as, or larger than, the reduction in computational cost from $\mathcal{O}(nC)$ to $\mathcal{O}(mC)$, so that the savings from subsampling are dominated by overhead cost. In contrast, when each contribution is expensive to evaluate, as in the aliased spectral setting considered here, the realised CPU gain is closer to the ideal ratio $n/m$.

As discussed in the introduction, the debiased Whittle accounts for aliasing and spectral blurring by replacing the spectral density with the expected periodogram. This involves a convolution with the Fej\'{e}r kernel, which can be computed efficiently using the FFT \citep{sykulski2019debiased}. While the resulting log-likelihood can still be written as a sum, as in \eqref{eq:Whittle_decomposed}, the corresponding terms $\ell_k(\boldsymbol{\theta})$ in the debiased Whittle depend on the full autocovariance sequence $\gamma(\tau;\boldsymbol{\theta})$ for $\tau = 0, \dots, n-1$; see \cite{sykulski2019debiased} for details. Consequently, although subsampling could in principle be applied, it does not lead to computational savings for the debiased Whittle likelihood. In particular, in the present setting with $C=\mathcal{O}(K)$, subsampling the standard Whittle likelihood is advantageous relative to the full-data debiased Whittle likelihood when
$$\mathcal{O}(mK) \ll \mathcal{O}(n \log n).$$
If we assume that the truncation level $K$ is fixed independently of $n$, so that $K=\mathcal{O}(1)$, this condition reduces to $\mathcal{O}(m)\ll \mathcal{O}(n \log n)$, which is satisfied in the effective subsampling regime $m \ll n$.

To summarise, these considerations identify a computational regime in which subsampling is particularly effective, namely when the number of likelihood contributions is large and the cost per contribution is substantial. In contrast, when contributions are inexpensive, the reduction from $n$ to $m$ yields limited computational gains. Furthermore, in settings where aliasing can be addressed through the debiased Whittle likelihood, subsampling may still be beneficial when evaluating the aliased spectral density for a small number of terms is computationally cheaper than computing the debiased Whittle likelihood, which requires FFT-type calculations over the full frequency spectrum for each value of $\boldsymbol{\theta}$.

\section{Experiments}\label{sec:experiments}

\subsection{L\'{e}vy-driven CARMA models}\label{subsec:models_experiments}
We consider continuous-time autoregressive moving average (CARMA) models driven by L\'{e}vy processes, observed at discrete time points; see Appendix \ref{app:carma_time_domain} for the corresponding continuous-time formulation. In the frequency domain, the underlying continuous-time CARMA($p,q$) process admits a spectral density of the form
\begin{align}\label{eq:continuous_time_spectral}
    f(\omega; \boldsymbol{\theta})
=
\frac{\sigma^2}{2\pi}
\left|
\frac{\beta(\mathrm{i}\omega)}
{\alpha(\mathrm{i}\omega)}
\right|^2, \,\, \omega\in\mathbb{R},
\end{align}
where 
\begin{align*}
    \alpha(z) &=
z^p + \alpha_1 z^{p-1} + \dots + \alpha_p, \quad
\beta(z)
=
1 + \beta_1 z + \dots + \beta_q z^q,
\end{align*}
denote the autoregressive and moving average polynomials, respectively, with the standard CARMA restriction $p>q$, and
\begin{align*}
    \boldsymbol{\theta} & = (\alpha_1, \dots, \alpha_p, \beta_1, \dots, \beta_q, \sigma^2).
\end{align*}
Stationarity and invertibility are enforced through priors on a transformed parameter space. For stationarity, the autoregressive polynomial is required to have roots with strictly negative real parts. Following \cite{tomasson2015some}, we assign a uniform prior over the corresponding region of the transformed space and use the Cayley-Hamilton transform \citep{belcher1994parameterization} to convert stationary AR$(p)$ parameters to continuous-time AR($p$) coefficients. For the moving average component, all our experiments in this section satisfy $q\leq 1$, so invertibility is enforced by assigning a uniform prior on $(-1,1)$. Finally, the variance parameter is assigned the prior
$$\log\sigma^2\sim \mathcal{N}(0,1).$$

\begin{figure}[h]
    \centering
    \includegraphics[width=0.95\linewidth]{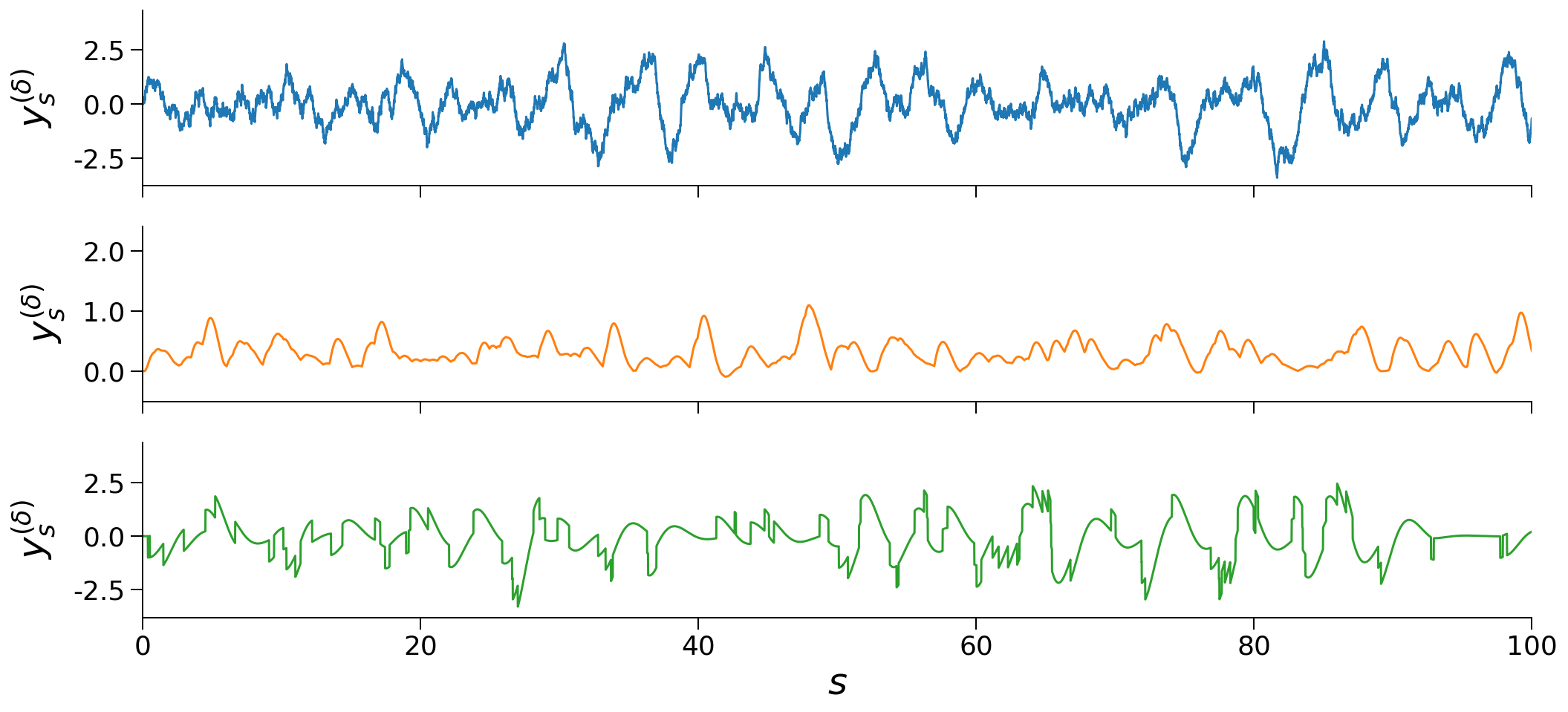}
    \caption{Simulated discrete time observations for the models in Section \ref{subsec:models_experiments}. The first 100 observations are shown. Rows correspond to Gaussian-driven CARMA$(2,1)$ (top), gamma-driven CARMA$(2,0)$ (middle), and two-sided Poisson-driven CARMA$(2,1)$ (bottom) models.}
    \label{fig:carma_sims}
\end{figure}

We consider simulation settings similar to those in \citet{fechner2018limit} through three examples, including Gaussian and two-sided Poisson-driven CARMA$(2,1)$ models, together with a gamma-driven CARMA$(2,0)$ model \citep{graf2009parametric}. Figure \ref{fig:carma_sims} shows representative simulated trajectories from each model. These specifications provide a range of departures from Gaussianity in the time-domain through continuous-time dynamics driven by different L\'evy processes, while admitting the common explicit spectral representations in \eqref{eq:continuous_time_spectral}. This reflects the fact that the spectral density characterises the second-order dependence structure of the process. In the Gaussian case driven by Brownian motion, this fully determines the dependence structure, whereas non-Gaussian L\'{e}vy-driven models additionally involve higher-order distributional properties that are not captured by the spectral density. Consequently, for finite samples, the degree of departure from Gaussianity is expected to influence the accuracy of the Whittle likelihood approximation to the exact time-domain likelihood, although the approximation improves asymptotically under standard regularity conditions \citep{CLT_FT, Asymptotic_properties_periodogram}.

To accurately approximate the underlying continuous-time process, simulations are performed on a fine time grid before retaining observations on the coarser sampling grid with spacing $\delta$. To form the corresponding discrete-time spectral density, sampling the process at spacing $\delta$ folds the continuous-time spectral density in \eqref{eq:continuous_time_spectral}, defined on $\mathbb{R}$, onto the interval $[-\pi/\delta,\pi/\delta]$ through the aliasing representation in \eqref{eq:aliased_spectrum}. Consequently, the severity of aliasing depends on how much spectral mass of the continuous-time process lies outside the Nyquist interval $[-\pi/\delta,\pi/\delta]$. When $\delta$ is sufficiently small, most of the spectral mass is concentrated within this interval and the contributions from further frequency shifts in the aliasing sum decay rapidly. In this setting, truncating the infinite sum in \eqref{eq:aliased_spectrum} to a small value of $K$ provides an accurate approximation to the aliased spectrum. Conversely, when substantial spectral mass lies outside the Nyquist interval, larger truncation levels are required to accurately capture the aliasing contributions. 

The choice of $\delta$ in the simulations below is therefore adapted to the goals of each experiment. In validating the Whittle likelihood assumptions and posterior accuracy experiments in Sections \ref{subsec:validation_experiment} and \ref{subsec:posterior_accuracy_experiment}, respectively, $\delta$ is chosen sufficiently small so that the discretely observed process closely reflects the underlying continuous-time model while requiring only a small truncation level $K$. In contrast, when studying the effect of aliasing truncation in Section \ref{subsec:aliasing_truncation_experiment}, we deliberately choose larger values of $\delta$ to induce more pronounced aliasing effects and increase the sensitivity to small values of $K$. It is worth noting that, for real data applications, the observation spacing $\delta$ is fixed by the data collection process and may induce substantial aliasing. In such settings, one cannot generally assume that a small truncation level $K$ provides an accurate approximation in \eqref{eq:aliased_spectrum}. In practice, larger values of $K$ may therefore be chosen conservatively to guard against non-negligible truncation error in the aliasing approximation, which motivates our computational regime of interest in Section \ref{subsec:computational_regimes}.

\subsection{Experiment 1: Validating the Whittle likelihood assumptions}\label{subsec:validation_experiment}

As briefly discussed in Section \ref{eq:model_and_likelihood}, the Whittle likelihood relies on the asymptotic properties of the periodogram ordinates, which are computed from the complex-valued discrete DFT in \eqref{eq:DFT}. Under standard regularity conditions, the DFT is asymptotically complex normal, with real and imaginary parts asymptotically independent, and the corresponding periodogram ordinates are exponentially distributed as in \eqref{eq:periodogram_dist} as the observation horizon $T\to\infty$ for fixed sampling interval $\delta$ (which implies $N\to \infty$).

We assess the finite-sample accuracy of these approximations for discretely observed realisations from the models in Section \ref{subsec:models_experiments}. For each specification, we simulate $2{,}000$ independent realisations of the observed process over a fixed time horizon $T=\delta (N-1)=1{,}000$, with observation spacing $\delta = 0.1$, yielding $N = 10{,}001$ time-domain observations and $n=(N-1)/2=5{,}000$ observations in the Whittle likelihood. The DFT and periodogram are evaluated at selected Fourier frequencies $k\in\{1, 10, 1000, 4000\}$, spanning low, moderate, and high frequencies in the spectrum. To approximate the aliased spectral density, the infinite sum in \eqref{eq:aliased_spectrum} is truncated to $|j|\leq K$, where $K=50$ is chosen sufficiently large so that the truncation error is negligible in this experiment.

\begin{figure}[h!]
    \centering
    \includegraphics[width=0.95\linewidth]{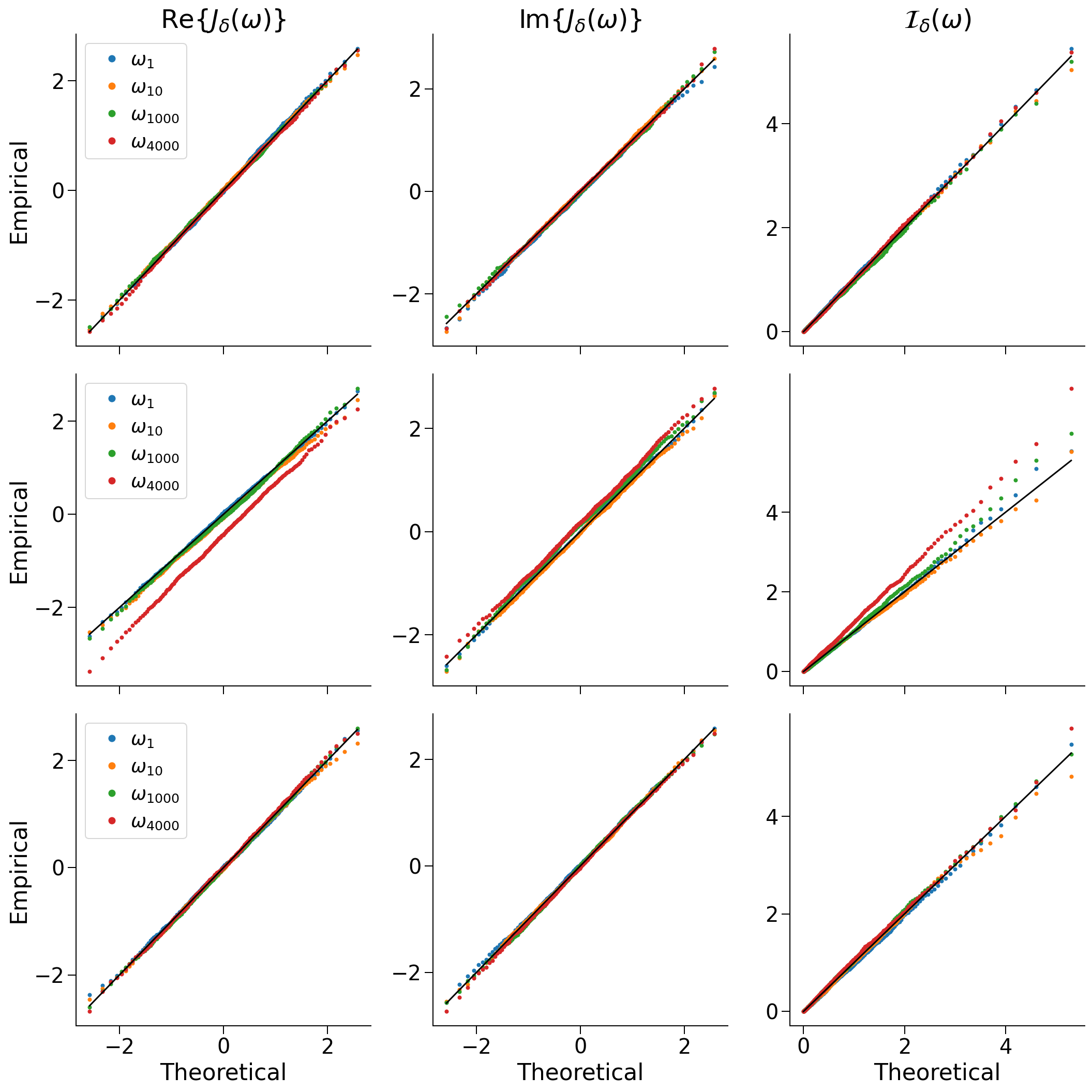}
    \caption{Results for Experiment 1 in Section \ref{subsec:validation_experiment}. Quantile-quantile plots for the standardised real and imaginary parts of the discrete Fourier transform (DFT) $J_\delta(\omega_k)$ and the periodogram ordinates $\mathcal{I}_\delta(\omega_k)=|J_\delta(\omega_k)|^2$ for the models in Section \ref{subsec:models_experiments}. The first two columns compare the standardised DFT components against the standard normal distribution, while the third column compares the periodogram against the exponential distribution in \eqref{eq:periodogram_dist}. Rows correspond to Gaussian-driven CARMA$(2,1)$ (top), gamma-driven CARMA$(2,0)$ (middle), and two-sided Poisson-driven CARMA$(2,1)$ (bottom) models.}
    \label{fig: QQ plots}
\end{figure}

Figure \ref{fig: QQ plots} shows the corresponding quantile-quantile (QQ) plots for the real and imaginary parts of the DFT together with the periodogram ordinates. The standardised real and imaginary DFT components show strong agreement with the standard normal reference distribution for the Gaussian-driven model, with reasonable agreement for the non-Gaussian L{\'e}vy-driven models despite some visible departures, particularly in the gamma-driven case at higher frequencies. The periodogram ordinates similarly show reasonable agreement with the exponential reference distribution in \eqref{eq:periodogram_dist}, although deviations are again more pronounced in the non-Gaussian models. Overall, these results provide empirical support for the Whittle likelihood approximation, while illustrating that finite-sample departures may be more noticeable in non-Gaussian settings.

\newpage
\subsection{Experiment 2: Posterior accuracy and CPU efficiency}\label{subsec:posterior_accuracy_experiment}

We study two posterior approximation comparisons of interest. The first concerns how well the full-data Whittle posterior approximates the posterior based on the exact discrete-time Gaussian likelihood in the time domain computed via filtering. This posterior corresponds to inference for the discretely observed continuous-time model and, for sufficiently small spacing $\delta$, provides an increasingly accurate approximation to inference based on the underlying continuous-time process. The second concerns how well the subsampling Whittle posterior approximates the corresponding full-data Whittle posterior. Motivated by the finite-sample deviations observed in Experiment 1, we simulate substantially longer time series in this experiment. This also places the analysis in a regime where subsampling is practically relevant. Specifically,  we use a fixed time horizon $T=20{,}000$ with observation spacing $\delta = 0.1$, yielding $N = T/\delta + 1 = 200{,}001$ time-domain observations, and hence $n=(N-1)/2 = 100{,}000$ observations in the Whittle likelihood. 

For the first comparison, concerning the time-domain posterior, we consider the model driven by Brownian motion, as the corresponding discretely observed CARMA process admits an exact Gaussian state-space representation. The parameters $\boldsymbol{\theta}$ can therefore be inferred using standard Markov chain Monte Carlo with the time-domain likelihood evaluated via the Kalman filter \citep{tsai2000note,TimeSeriesBook_Harvey, jones1981fitting}. In contrast, state-space formulations for the other two L\'evy-driven models require particle filter approximations to the filtering and are therefore excluded from this comparison. 

For the second comparison, concerning subsampling, we follow \citet{salomone2020spectral}, who propose a grouped version of \eqref{eq:diff_estimator}, in which the log-likelihood contributions are partitioned into $G$ groups and control variates are constructed at the group level rather than for individual contributions. \citet{salomone2020spectral} argue that constructing a quadratic approximation for sums of contributions is more effective in this setting, as the Bernstein--von Mises theorem implies that the log-likelihood is approximately quadratic in a neighbourhood of the posterior mode for sufficiently large samples. We implement subsampling MCMC with $G=1{,}000$ groups and subsample size $m=10$, corresponding to $1\%$ of the grouped likelihood contributions per iteration. 

\begin{figure}[h]
    \centering
    \includegraphics[width=0.95\linewidth]{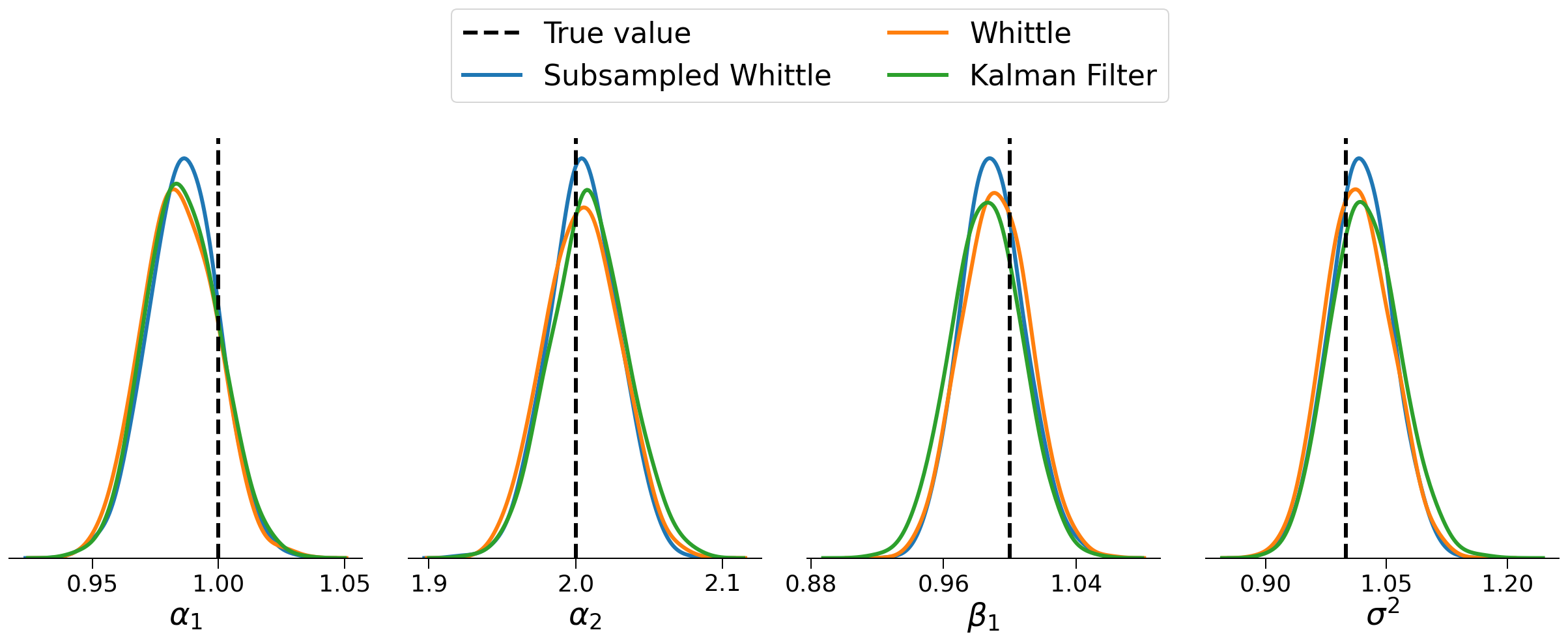}
    \caption{Results for Experiment 2 in Section \ref{subsec:posterior_accuracy_experiment}. Posterior marginal distributions obtained from full-data MCMC and subsampling MCMC for the Gaussian-driven CARMA$(2,1)$ model fitted to simulated data. Vertical dashed lines indicate the true parameter values.}
    \label{fig:gaussian_21_kde_posteriors}
\end{figure}

Figure \ref{fig:gaussian_21_kde_posteriors} shows that the subsampled Whittle and full-data Whittle posteriors are in close agreement for the Gaussian-driven CARMA model. Moreover, the figure shows that the full-data time-domain posterior is also close to the full-data Whittle posterior. 

\begin{figure}[h]
    \centering
    \includegraphics[width=0.95\linewidth]{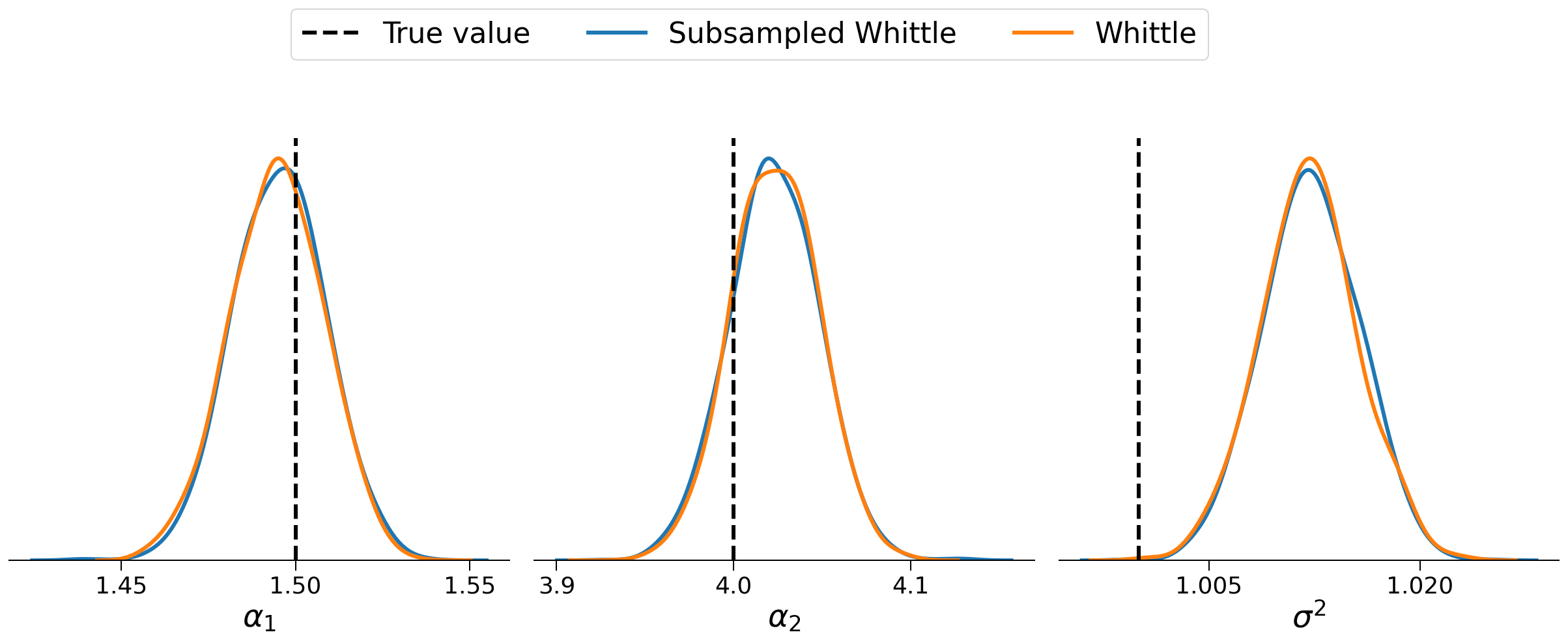}
    \caption{Results for Experiment 2 in Section \ref{subsec:posterior_accuracy_experiment}. 
    Posterior marginal distributions obtained from full-data MCMC and subsampling MCMC for the gamma CARMA$(2,0)$ model fitted to simulated data. Vertical dashed lines indicate the true parameter values.}
    \label{fig:gamma_20_kde_posteriors}
\end{figure}

\begin{figure}[h]
    \centering
    \includegraphics[width=0.95\linewidth]{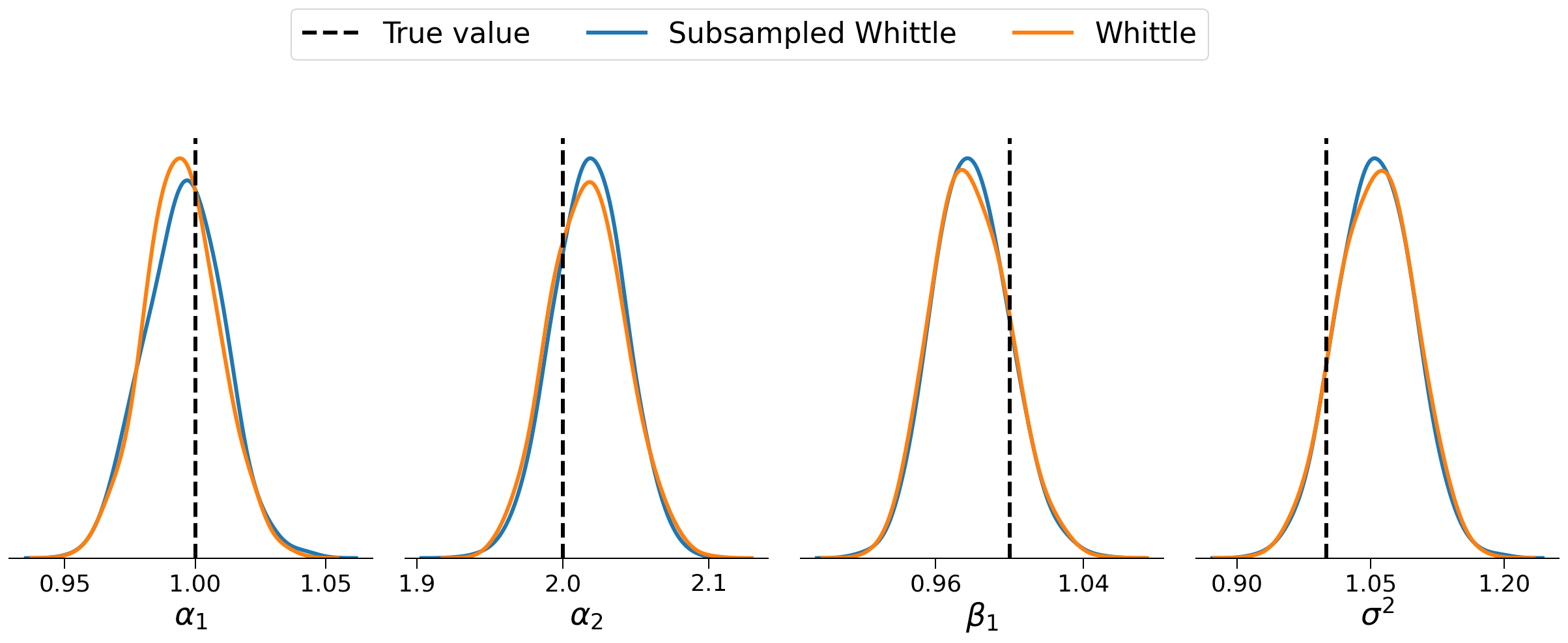}
    \caption{Results for Experiment 2 in Section \ref{subsec:posterior_accuracy_experiment}. 
    Posterior marginal distributions obtained from full-data MCMC and subsampling MCMC for the two-sided Poisson CARMA$(2,1)$ model fitted to simulated data. Vertical dashed lines indicate the true parameter values.}
    \label{fig:poisson_21_kde_posteriors}
\end{figure}

Figures \ref{fig:gamma_20_kde_posteriors} and \ref{fig:poisson_21_kde_posteriors} show similarly strong agreement between the subsampled and full-data Whittle posteriors for the gamma- and Poisson-driven models, respectively. No time-domain posteriors are shown for these non-Gaussian models, as they are not readily available computationally, as discussed above.

Having established posterior accuracy, we now turn to computational efficiency. Using $1\%$ of the grouped likelihood contributions at each iteration, the subsampling scheme yields a nominal $100\times$ reduction in computational cost relative to evaluating the full-data likelihood, ignoring overhead costs. Previous subsampling MCMC work reports nominal computational gains based on the reduction in likelihood contribution evaluations \citep{quiroz2018delay,quiroz2019speeding,dang2019hamiltonian,salomone2020spectral,villani2024spectral}. In the present setting, where individual likelihood contributions are computationally expensive due to aliased spectral density evaluation, we are interested in whether realised CPU gains can approach these nominal reductions. Table \ref{tab:cpu_speedups} shows the realised CPU speed-ups for the models in Section \ref{subsec:models_experiments}.

\begin{table}[H]
\centering
\caption{Nominal and realised CPU speed-ups for subsampling MCMC using a 1\% subsample in the simulation experiments. The nominal speed-up assumes computational cost scales exactly with the number of likelihood contributions.}
\label{tab:cpu_speedups}
\begin{tabular}{lcc}
\toprule
Model & Nominal speed-up & Realised CPU speed-up \\
\midrule
Gaussian-driven CARMA$(2,1)$          & 100.0 & 83.5 \\
Gamma-driven CARMA$(2,0)$             & 100.0 & 85.6 \\
Two-sided Poisson-driven CARMA$(2,1)$ & 100.0 & 86.9 \\
\bottomrule
\end{tabular}
\end{table}

Table \ref{tab:cpu_speedups} shows realised CPU speed-ups of approximately $84\times$ to $87\times$, demonstrating that realised CPU gains can closely approach nominal subsampling gains when likelihood contributions are computationally expensive.

\subsection{Experiment 3: Effect of aliasing truncation}\label{subsec:aliasing_truncation_experiment}
Since the Whittle likelihood depends directly on $f_\delta(\omega_k;\boldsymbol{\theta})$, the truncation error in the aliased spectral density propagates directly into the likelihood approximation. We therefore focus on the spectral approximation error itself, rather than separately studying posterior sensitivity to $K$. We study the effect of truncating the aliasing sum in \eqref{eq:aliased_spectrum} to $|j|\leq K$. The aim is to quantify when small values of $K$ provide accurate approximations to the aliased spectral density and when the truncation error becomes non-negligible.

For a fixed CARMA specification, we compute a high-accuracy reference approximation
\begin{align}\label{eq:aliased_spectrum_ref}
f^{\mathrm{ref}}_\delta(\omega; \boldsymbol{\theta})
= \sum_{|j| \leq K_{\mathrm{ref}}}
f\left(\omega + \frac{2\pi j}{\delta}; \boldsymbol{\theta}\right), \,\, \omega \in \left[-\frac{\pi}{\delta},\frac{\pi}{\delta}\right],
\end{align}
with $f$ in \eqref{eq:continuous_time_spectral}, here $K_{\mathrm{ref}}$ is chosen sufficiently large so that the truncation error is negligible. We then compare this to the truncated approximation
\begin{align}\label{eq:aliased_spectrum_trunc}
f^{K}_\delta(\omega; \boldsymbol{\theta})
= \sum_{|j| \leq K}
f\left(\omega + \frac{2\pi j}{\delta}; \boldsymbol{\theta}\right), \,\, \omega \in \left[-\frac{\pi}{\delta},\frac{\pi}{\delta}\right],
\end{align}
for increasing values of $K$ and several choices of the sampling interval $\delta$. Since evaluating the truncated aliased spectral density in \eqref{eq:aliased_spectrum_trunc} requires summing over $2K+1$ shifted frequencies, the per-frequency computational cost scales linearly with $K$. Consequently, larger truncation levels directly increase the cost of evaluating the Whittle likelihood, motivating the computational regime considered in the present work. We quantify the error when approximating \eqref{eq:aliased_spectrum_ref} with \eqref{eq:aliased_spectrum_trunc} using the relative integrated error, where the integration is approximated by summation over the Fourier frequencies, i.e.\
\begin{align}\label{eq:integrated_error}
    E_K(\delta)
& =
\frac{
\sum_{\omega_k}
\left|f_{\delta}^{(K)}(\omega_k;\boldsymbol{\theta})
-
f_{\delta}^{\mathrm{ref}}(\omega_k;\boldsymbol{\theta})\right|
}{
\sum_{\omega_k}
f_{\delta}^{\mathrm{ref}}(\omega_k;\boldsymbol{\theta})
}.
\end{align}

\begin{figure}[H]
    \centering
    \includegraphics[width=0.92\linewidth]{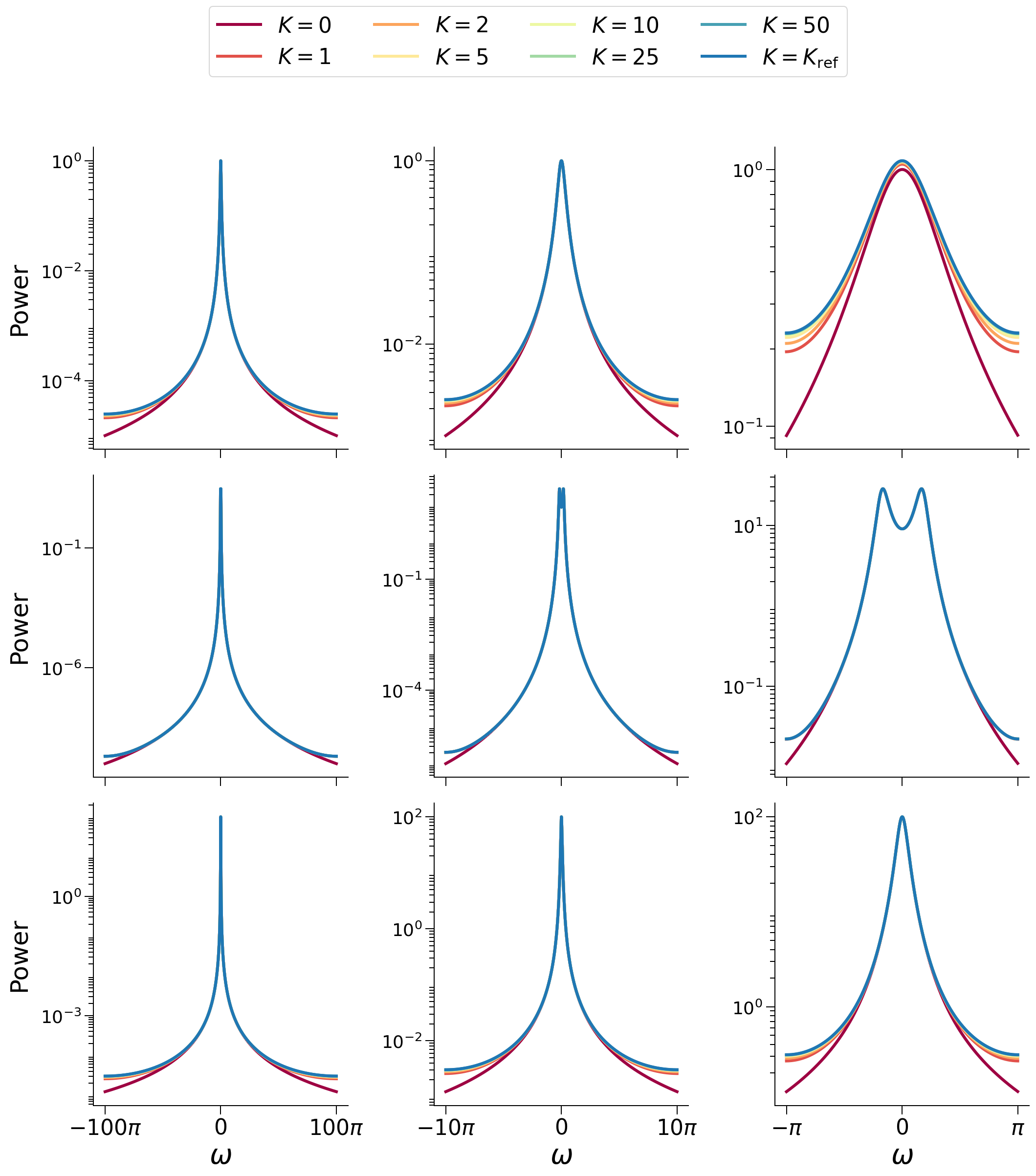}
    \caption{Results for Experiment 3 in Section \ref{subsec:aliasing_truncation_experiment}. Aliased spectral density approximations on a logarithmic scale for the models in Section \ref{subsec:models_experiments}. Each panel shows the aliased spectral density approximation for increasing truncation levels $K$ (see the legend), together with the reference approximation using $K_{\mathrm{ref}}=500$. Rows correspond to Gaussian-driven CARMA$(2,1)$ (top), gamma-driven CARMA$(2,0)$ (middle), and two-sided Poisson-driven CARMA$(2,1)$ (bottom) models. Columns correspond to sampling intervals $\delta = 0.01, 0.1, 1$ (left to right).}
    \label{fig:aliased_spectra_approximations}
\end{figure}

Figure \ref{fig:aliased_spectra_approximations} illustrates how small truncation levels may distort the aliased spectral density when aliasing effects are non-negligible. For small sampling intervals, the aliased spectral density is well approximated even for modest truncation levels, reflecting the relative weak contribution from higher-frequency aliases. In contrast, as the sampling interval increases, aliasing becomes more pronounced and substantially larger truncation levels are required to accurately approximate the spectrum.

\begin{figure}[H]
    \centering
    \includegraphics[width=0.95\linewidth]{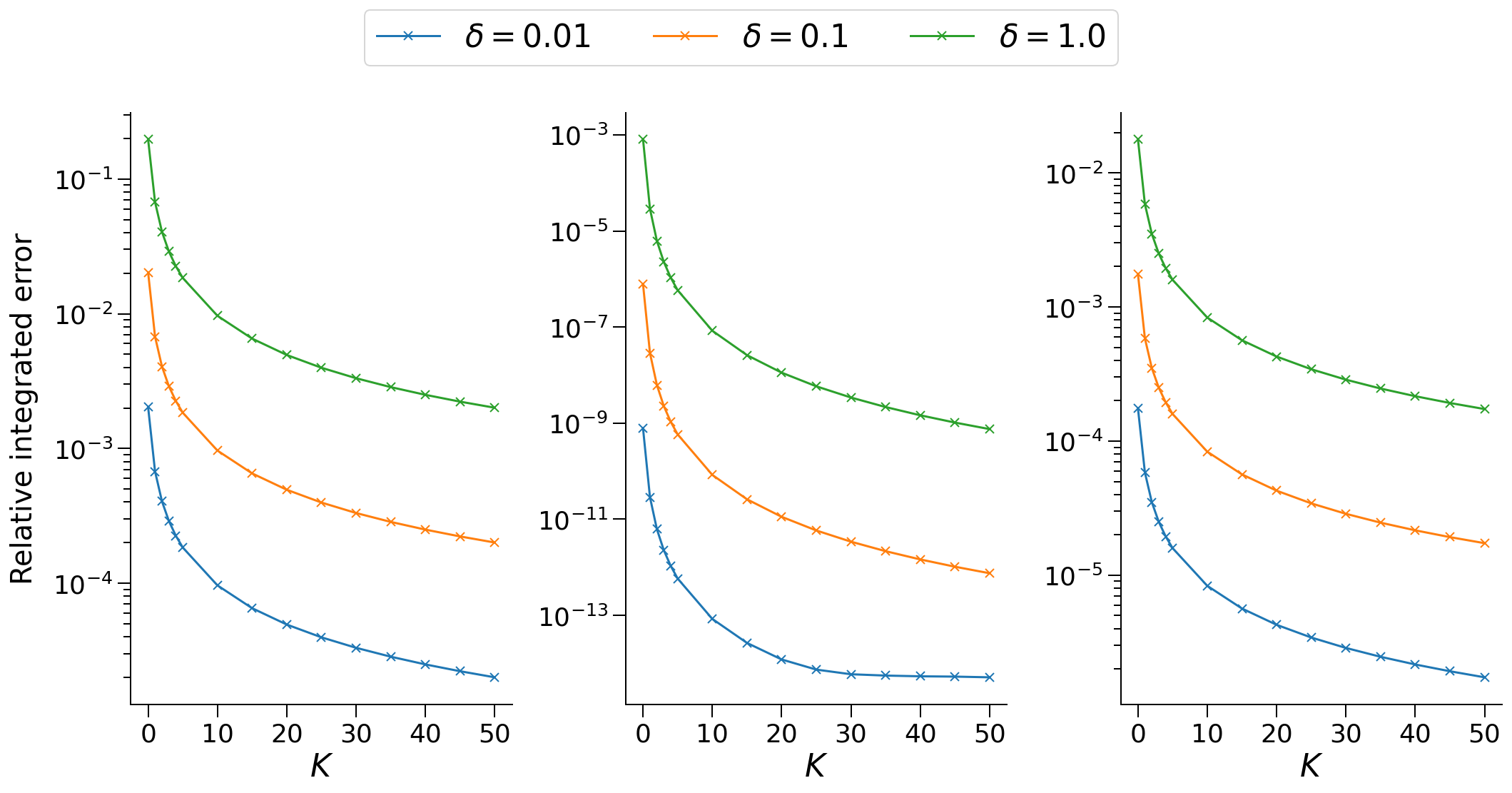}
    \caption{Results for Experiment 3 in Section \ref{subsec:aliasing_truncation_experiment}. Log-scale integrated approximation error, defined in \eqref{eq:integrated_error},  of the truncated aliased spectral density as a function of the truncation level $K$ for the models in Section \ref{subsec:models_experiments}. Panels correspond to Gaussian-driven CARMA$(2,1)$ (left), gamma-driven CARMA$(2,0)$ (middle), and two-sided Poisson-driven CARMA$(2,1)$ (right) models. Curves correspond to sampling intervals $\delta = 0.01$, $0.1$, and $1$.}
\label{fig:spectra_approximation_integrated_error}
\end{figure}

Figure \ref{fig:spectra_approximation_integrated_error} complements these observations by showing the relative integrated error in \eqref{eq:integrated_error} as a function of $K$ for different values of $\delta$ on a logarithmic scale. The approximation error decreases monotonically as the truncation level increases, but the required truncation level depends strongly on the sampling interval due to the severity of aliasing. These results reveal a computational regime in which subsampling methods can be particularly effective, as strongly aliased settings require expensive per-frequency spectral evaluations, and subsampling reduces the number of frequencies that must be evaluated.

\section{Application}\label{sec:application}

To demonstrate the methodology in a real-data setting, we consider high-frequency bitcoin data consisting of $N=200{,}001$ observations ($n=100{,}000$ Whittle likelihood contributions) of log-squared returns, observed at one-minute intervals, obtained from Coinbase\footnote{The data can be downloaded from \url{http://www.coinbase.com}.}. We therefore set the sampling interval to $\delta=1$, so that the observable spectral domain is $[-\pi, \pi]$. As the Whittle likelihood depends only on the spectral density, we do not need to specify a particular L{\'e}vy driving process for the application. To provide a realistic application setting, we first perform a small model selection exercise over a few low-order CARMA specifications, and then compare subsampling MCMC and full-data MCMC for the selected model. Stationarity and invertibility constraints are enforced as described in Section \ref{subsec:models_experiments}.

We fit CARMA$(p, q)$ models with $0\leq q<p\leq 2$ and select the model with the smallest Bayesian information criterion (BIC). To compute the BIC, the posterior mode is obtained using global optimisation via Basin-hopping \citep{li1987monte}. Table \ref{tab:CARMA_BIC_table} shows the resulting values, with CARMA$(2,1)$ selected according to the BIC.

\begin{table}[h]
\centering
\caption{Bayesian information criterion (BIC) values for candidate CARMA specifications fitted to the bitcoin data in Section \ref{sec:application}. Smaller values indicate a better fit; the smallest value is shown in bold.}
\label{tab:CARMA_BIC_table}
\begin{tabular}{lcc}
\toprule
 & $q=0$ & $q=1$ \\
\midrule
CARMA$(1,q)$ & 1,623,423 & -- \\
CARMA$(2,q)$ & 7,305,523 & \textbf{1,607,877} \\
\bottomrule
\end{tabular}
\end{table}

\begin{figure}[h]
    \centering
    \includegraphics[width=0.95\linewidth]{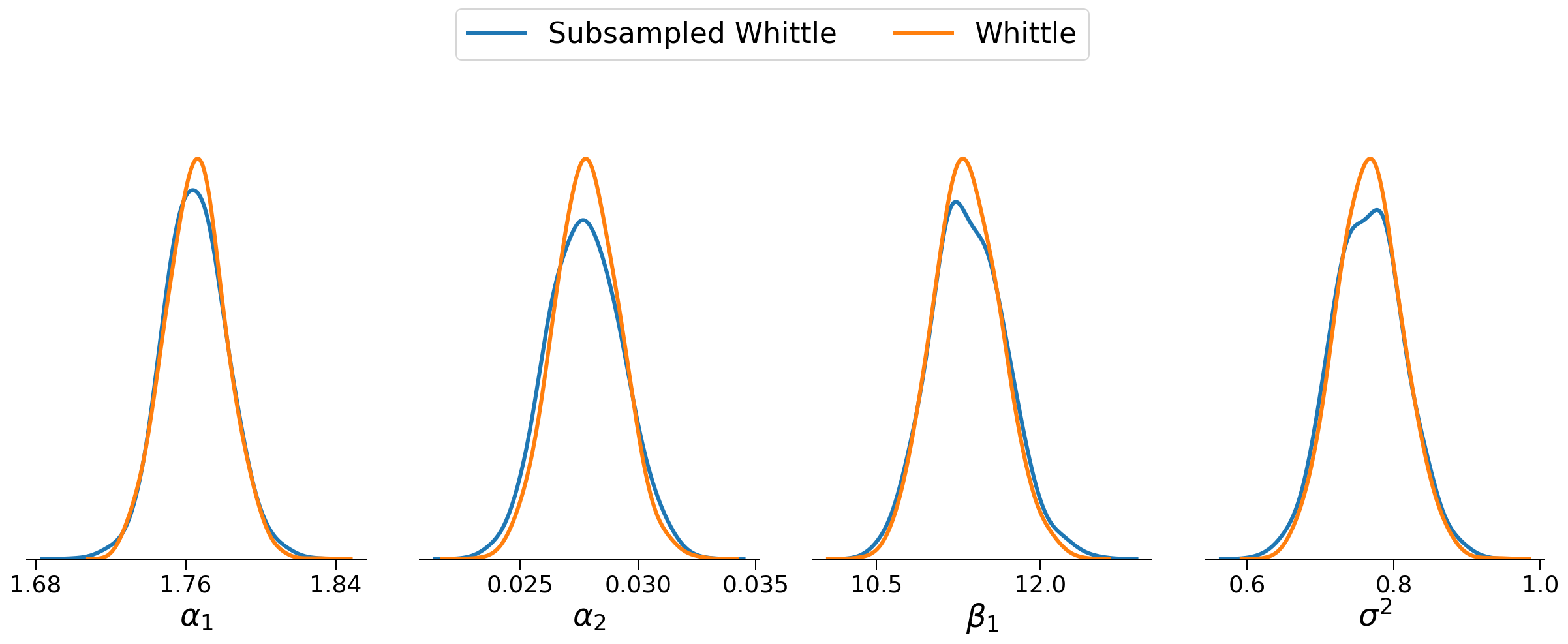}
    \caption{Results for the application in Section \ref{sec:application}. Posterior marginal distributions obtained from full-data MCMC and subsampling MCMC for the CARMA$(2,1)$ model fitted to bitcoin data.}
    \label{fig:bitcoin_21_kde_posteriors}
\end{figure}

Figure \ref{fig:bitcoin_21_kde_posteriors} shows kernel density estimates of the posterior marginals for the preferred CARMA$(2,1)$ specification under subsampling MCMC and full-data MCMC inference. The two posterior distributions are nearly indistinguishable, indicating that subsampling preserves posterior accuracy in this application.

Finally, to assess sampling efficiency after accounting for runtime, we combine integrated autocorrelation time (IACT) and CPU time through the computational time (CT),
\begin{align}\label{eq:CT_def_app}
    \mathrm{CT} = \mathrm{IACT} \times \mathrm{CPU},
\end{align}
with smaller values indicating greater efficiency. We then define the relative computational time (RCT) for subsampling MCMC (sub) relative to full-data MCMC (full) as
$$
\mathrm{RCT} = \frac{\mathrm{CT}_{\mathrm{full}}}{\mathrm{CT}_{\mathrm{sub}}},
$$
so that $\mathrm{RCT}>1$ indicates that subsampling is more efficient overall. Table \ref{tab:bitcoin_efficiency} shows the corresponding IACT and CPU values, from which the RCT values range from $61\times$ to $76\times$ across parameters.

\begin{table}[h]
\centering
\caption{Results for the bitcoin application. Sampling efficiency is measured using integrated autocorrelation time (IACT). Central processing unit (CPU) time is reported in seconds. Both methods run for $13{,}000$ Markov chain Monte Carlo iterations, including $3{,}000$ burn-in iterations.}
\label{tab:bitcoin_efficiency}
\begin{tabular}{lccccccc}
\toprule
& \multicolumn{4}{c}{IACT} & & \\
\cmidrule(lr){2-5}
Method & $\alpha_1$ & $\alpha_2$ & $\beta_1$ & $\sigma^2$ & CPU  & Realised CPU speed-up \\
\midrule
Full-data Whittle   & 21.1 & 16.3 & 18.8 & 19.3 & 6293.4 & $1.0\times$ \\
Subsampling Whittle & 25.2 & 24.2 & 24.2 & 24.0 & 69.5   & $90.5\times$ \\
\bottomrule
\end{tabular}
\end{table}

These results demonstrate that subsampling MCMC can deliver substantial overall efficiency gains in this application without compromising posterior accuracy.

\section{Conclusion and future research}\label{sec:conclusion_and_future_research}

We study subsampling-based Bayesian frequency-domain inference for continuous-time models observed at equally spaced discrete time points, with computationally expensive likelihood contributions arising from aliasing, and characterise the computational regime in which genuine CPU gains from subsampling are achievable in practice. Through simulation experiments and a real-data application to high-frequency bitcoin data, we show that the subsampled Whittle posterior closely approximates the corresponding full-data Whittle posterior while delivering substantial computational gains.

A natural direction for future research is the development of subsampling strategies for debiased Whittle likelihoods, which could further broaden the applicability of scalable Bayesian frequency-domain inference. Finally, although our paper focuses on frequency-domain inference, the time-domain comparison considered in Section \ref{subsec:posterior_accuracy_experiment} also suggests an interesting open problem. Exact likelihood evaluation in the Gaussian case relies on specialised continuous-time state-space formulations and recursive filtering methodology for discretely observed continuous-time processes. To the best of our knowledge, it remains unclear whether related ideas can be extended to conditionally Gaussian continuous-time models, including settings with unequally spaced observations \citep{jones1987continuous}, potentially through methods such as the mixture Kalman filter in \cite{chen2000mixture}.
\newpage

\appendix
 
\section{Continuous-time ARMA models in the time domain}
\label{app:carma_time_domain}

Our methodology is formulated in the frequency domain, where inference is based on the spectral density of the observed process. This spectrum is obtained through aliasing of the underlying continuous-time spectral density, given by the Fourier transform of the continuous autocovariance function. For readers interested in the underlying dynamics of the process, this appendix briefly outlines the corresponding time-domain formulation of the L{\'e}vy-driven CARMA models considered in the paper. See \cite{applebaum2009levy} for a textbook treatment of L{\'e}vy processes in general, and \cite{brockwell2001levy, brockwell2004representations} for their use in continuous-time autoregressive moving average models in particular.

A continuous-time autoregressive moving average CARMA$(p,q)$ process with $p>q$ is driven by a L{\'e}vy process $\{L(t)\}_{t\geq 0}$, that is, a stochastic process with stationary and independent increments encompassing Brownian motion as well as jump processes. Its formal stochastic differential equation representation is given by
\begin{align}\label{eq:Formal_SDE_Levy}
    \alpha(D)Y(t) = \beta(D)DL(t),
\end{align}
where $D$ denotes the differentiation operator with respect to time, and the autoregressive and moving average polynomials are given by
\begin{align*}
\alpha(z) &= z^p + \alpha_1 z^{p-1} + \dots + \alpha_p, \\
\beta(z) &= 1 + \beta_1 z + \dots + \beta_q z^q.
\end{align*}

The L{\'e}vy-driven formulation extends the classical Brownian-motion-driven Gaussian setting by allowing more general driving noise processes capable of exhibiting features such as jumps, asymmetry, and heavy-tailed behaviour. Since L{\'e}vy processes are generally nowhere differentiable, the formal representation in \eqref{eq:Formal_SDE_Levy} should instead be understood through the following state-space stochastic differential equation,
\begin{align*}
Y(t) &= \boldsymbol{\beta}^{\top}\mathbf{X}(t), \\
d\mathbf{X}(t) &= \mathbf{A}\mathbf{X}(t)\,dt + \mathbf{R}\,dL(t),
\end{align*}
where
\begin{align*}
\mathbf{A} &=
\begin{bmatrix}
0 & 1 & 0 & \cdots & 0 \\
0 & 0 & 1 & \cdots & 0 \\
\vdots & \vdots & \vdots & \ddots & \vdots \\
0 & 0 & 0 & \cdots & 1 \\
-\alpha_p & -\alpha_{p-1} & -\alpha_{p-2} & \cdots & -\alpha_1
\end{bmatrix},
&
\mathbf{R} &=
\begin{bmatrix}
0 \\
0 \\
\vdots \\
0 \\
1
\end{bmatrix},
\end{align*}
and
\begin{align*}
\boldsymbol{\beta} =
\begin{bmatrix}
1 \\
\beta_1 \\
\vdots \\
\beta_q \\
\mathbf{0}
\end{bmatrix},
\end{align*}
with $\mathbf{0}$ denoting a vector of $(p-q-1)$ zeros.

Under standard stationarity conditions, namely that the roots of $\alpha(z)$ have strictly negative real parts, the process admits a stationary solution. Its second-order structure determines the autocovariance function of the stationary process, whose Fourier transform yields the underlying continuous-time spectral density used in the construction of the aliased observed-process spectrum in the main text.


\bibliographystyle{apalike}
\addcontentsline{toc}{section}{\refname}
\bibliography{ref}

\end{document}